\documentclass[10pt,twocolumn,superscriptaddress,floats,showpacs,nobalancelastpage,longbibliography,prb]{revtex4-2}

\usepackage[pdftex]{color}
\usepackage[dvipsnames]{xcolor}
\usepackage{braket}

\usepackage{array,longtable}
\newcolumntype{L}{>{\tiny $}p{0.33\columnwidth}<{$}}
\newcolumntype{M}{>{\scriptsize $}p{0.33\columnwidth}<{$}}
\newcolumntype{N}{>{\scriptsize $}p{0.43\columnwidth}<{$}}
\usepackage{amsmath}
\usepackage{amssymb}
\usepackage{amsfonts}
\usepackage{graphicx}
\usepackage{hyperref}
\usepackage{tabularx}
\usepackage{wasysym}
\usepackage{bm}

\usepackage{float}
\usepackage{graphics}
\usepackage[caption=false]{subfig}
\usepackage{tikz}
\usepackage{times}
\usepackage{dsfont}
\usepackage{subfig}

\usepackage[normalem]{ulem}

\allowdisplaybreaks[1]

\newif\ifhyper
\hypertrue
\ifhyper
\hypersetup{
   citecolor = {OliveGreen},
   colorlinks = {true}, 
   urlcolor = {RoyalBlue}, 
   linkcolor = {RoyalBlue}
}
\fi

\newenvironment{diagram}
{
\begin{tikzpicture}[baseline = (X.base),every node/.style={scale=0.8},scale=0.45]
}
{
\end{tikzpicture}
}

\begin{document}

\title{Probing the Dynamical Structure Factor of Quantum Spin Chains via Low-Temperature Gibbs States with Matrix Product State Subspace Expansion}

\author{Tomoya Takahashi}
\affiliation{Graduate School of Science and Technology, Keio University, Yokohama, Kanagawa 223-8522, Japan}

\author{Wei-Lin Tu}
\email{weilintu@keio.jp}
\affiliation{Graduate School of Science and Technology, Keio University, Yokohama, Kanagawa 223-8522, Japan}
\affiliation{Keio University Sustainable Quantum Artificial Intelligence Center (KSQAIC), Keio University, Tokyo 108-8345, Japan}

\author{Ji-Yao Chen}
\affiliation{Center for Neutron Science and Technology, Guangdong Provincial Key Laboratory of Magnetoelectric Physics and Devices, School of Physics, Sun Yat-sen University, Guangzhou 510275, China}

\author{Yusuke Nomura}
\affiliation{Institute for Materials Research (IMR), Tohoku University, Sendai 980-8577, Japan}
\affiliation{Advanced Institute for Materials Research (WPI-AIMR), Tohoku University, Sendai 980-8577, Japan}

\date{\today}

\begin{abstract}

Studying finite-temperature properties with tensor networks is notoriously difficult, especially at low temperatures, due to the rapid growth of entanglement and the complexity of thermal states. Existing methods like purification and minimally entangled typical thermal states offer partial solutions but struggle with scalability and accuracy in low-temperature regime. To overcome these limitations, we propose a new approach based on generating-function matrix product states (GFMPS). By directly computing a large set of Bloch-type excited states, we construct Gibbs states that moderate the area-law constraint, enabling accurate and efficient approximation of low-temperature thermal behavior. Our benchmark results show magnificent agreement with both exact diagonalization and experimental observations, validating the accuracy of our approach. This method offers a promising new direction for overcoming the longstanding challenges of studying low-temperature properties within the tensor network framework. We also expect that our method will facilitate the numerical simulation of quantum materials in comparison with experimental observations.

\end{abstract}

\maketitle

\section{Introduction}
\label{sec:Introduction}

Tensor network (TN) methods have become a cornerstone in the study of quantum many-body systems, achieving remarkable success over the past three decades~\cite{Orus2019, Cirac2021, Banuls2023}.
These advances have not only deepened theoretical insights but also connected computational approaches to experimental observations.
In condensed-matter physics, on the other hand, strongly correlated quantum systems often exhibit exotic phenomena at low temperatures, especially in low-dimensional settings where quantum effects are pronounced. 
TNs, particularly those where the entanglement entropy follows the area law, are now widely regarded as ideal tools for numerical investigations of such systems and their quantum ground states~\cite{White1992, Ostlund1995, Vidal2008}.

Beyond ground-state analysis, however, studying finite temperature properties using tensor network methods presents a formidable challenge due to the exponential growth of entanglement and the complexity of thermal states. 
Unlike ground states, which often obey the area law and can be efficiently represented by tensor networks, thermal states typically require capturing volume-law entanglement, making their representation far more demanding.
Techniques such as purification~\cite{Zwolak2004, Verstraete2004_2, Feiguin2005}, minimally entangled typical thermal states~(METTS)~\cite{White2009, Stoudenmire2010}, or exponential tensor renormalization group~(XTRG)~\cite{B.B.Chen2018} attempt to circumvent this obstacle with density matrix renormalization group (DMRG) or matrix product operator~(MPO), followed by meaningful applications~\cite{Chen2019, Li2019, Wietek2019, Jiménez2021}.
However, the above-mentioned methods might come with limitations in scalability due to the rapid entanglement growth. 
More importantly, both methods face significant challenges as the temperature approaches to zero, making it difficult to accurately capture the low-temperature behavior of quantum systems. This limitation stems from the rapid growth of entanglement and the increasing complexity of thermal states, which strain the efficiency and reliability of tensor network representations in this regime.
As a result, while progress has been made, finite temperature simulations with TNs are still constrained by algorithmic complexity and resource demands, spurring the development of new approaches.

On the other hand, the Gibbs state is known to characterize the equilibrium properties of a system at a given temperature. This is because it maximizes the von Neumann entropy subject to a constraint on the average energy~\cite{Jaynes1957}.
While it has been extensively investigated that Monte Carlo sampling on 
\begin{equation}
\rho_\text{G}=\frac{1}{Z}e^{-\beta \hat{H}}, 
\label{Gibbs_state}
\end{equation}
where $\beta$ is the inverse temperature and $Z$ represents the partition function, provides an useful tool for a classical solver of Gibbs state, there are still some limitations for low-temperature simulation (see, e.g., theorem 11 in Ref.~\onlinecite{Eldan2020}).
Recent studies have shown that quantum advantage in Gibbs state simulation diminishes once the temperature, or equivalently, the entropy, exceeds a certain threshold~\cite{Besserve2025}.
%
These findings suggest that simulating Gibbs states at low temperatures remains a fundamental challenge, regardless of whether quantum or classical methods are employed.

While the above-mentioned difficulties exist, nevertheless, numerical simulation of the dynamical structure factor is essential for interpreting and validating experimental observations in quantum materials, especially in strongly correlated systems.
The dynamical structure factor (DSF), $S(\textbf{k}, \omega)$, encodes the momentum- and energy-resolved response of a system to external probes such as neutrons or photons, making it a direct bridge between theory and experiment~\cite{Hansen2006}.
In quantum spin chains and other low-dimensional systems, experimental techniques like inelastic neutron scattering or resonant inelastic X-ray scattering provide high-resolution data on excitations~\cite{Ament2011}, but their interpretation hinges on accurate theoretical modeling.
Moreover, recent experiments conducted at extremely low temperatures can reveal pronounced quantum effects, making it essential to simulate DSF with the corresponding model Hamiltonian for meaningful comparison~\cite{L.S.Wu2019, Nikitin2021, Kish2025}.
However, simulating DSF is knowingly challenging due to the exponential complexity of quantum many-body dynamics and the need for real-time or frequency-domain resolution.
Especially, the computational difficulty becomes even more pronounced when evaluating the DSF at finite temperatures. In this regime, conventional DMRG-based approaches typically require performing two separate time evolutions, one for the purification of the thermal state and another for the dynamics. 
This doubled time‑evolution procedure significantly amplifies the growth of entanglement entropy, which in turn restricts the accessible simulation time and dramatically increases the computational cost. 
As a result, finite‑temperature DSF calculations often face severe limitations compared with their zero‑temperature counterparts.

In this work, we introduce a novel approach for investigating finite-temperature behavior, particularly in the low-temperature regime, based on our recently developed eigensolvers for excited states probed by the generating-function tensor network state (GFTNS)~\cite{Tu2021, Tu2024}. 
In contrast to recently proposed approaches that construct an effective Hamiltonian by projecting the original system onto its low-energy subspace and subsequently diagonalizing it~\cite{Cocchiarella2025}, our method directly solves for a large number of excited states, represented as Bloch states, and uses them to explicitly construct the finite-temperature Gibbs state within the low-energy eigenbasis.
Although tensor networks are normally constrained by the entanglement area law, our excited-state ansatz consists of structured summations over similar tensor diagrams. 
Such a summation of many tensor copies makes it possible to surpass the area-law limitation, as can be seen in our previous analysis for the Rényi entropy calculation~\cite{Tu2021}~(cf. thermal pure quantum matrix product state~\cite{Iwaki2021, Gohlke2023}). 
This enables a more expressive representation of low-energy excitations and allows our Gibbs states to serve as a reliable and direct approximation of thermal states at low temperatures.
While the present study focuses on one-dimensional (1D) systems, making use of the generating-function matrix product state (GFMPS), to establish the validity of our method, the underlying framework is readily extendable to higher-dimensional settings. 
Given that the computational complexity for obtaining plural excited states is comparable to that of ground states despite some overhead, our approach remains applicable provided that accurate ground-state synthesis is achievable. A comprehensive investigation of its performance in higher dimensions is then an important direction for our future research.

\section{Results and Discussion}
\label{sec:results}

\subsection{Gibbs state by MPS and DSF calculation}
\label{sec:MPS_Gibbs_state}

\begin{figure*}
  \centering
 \includegraphics[scale=0.55]{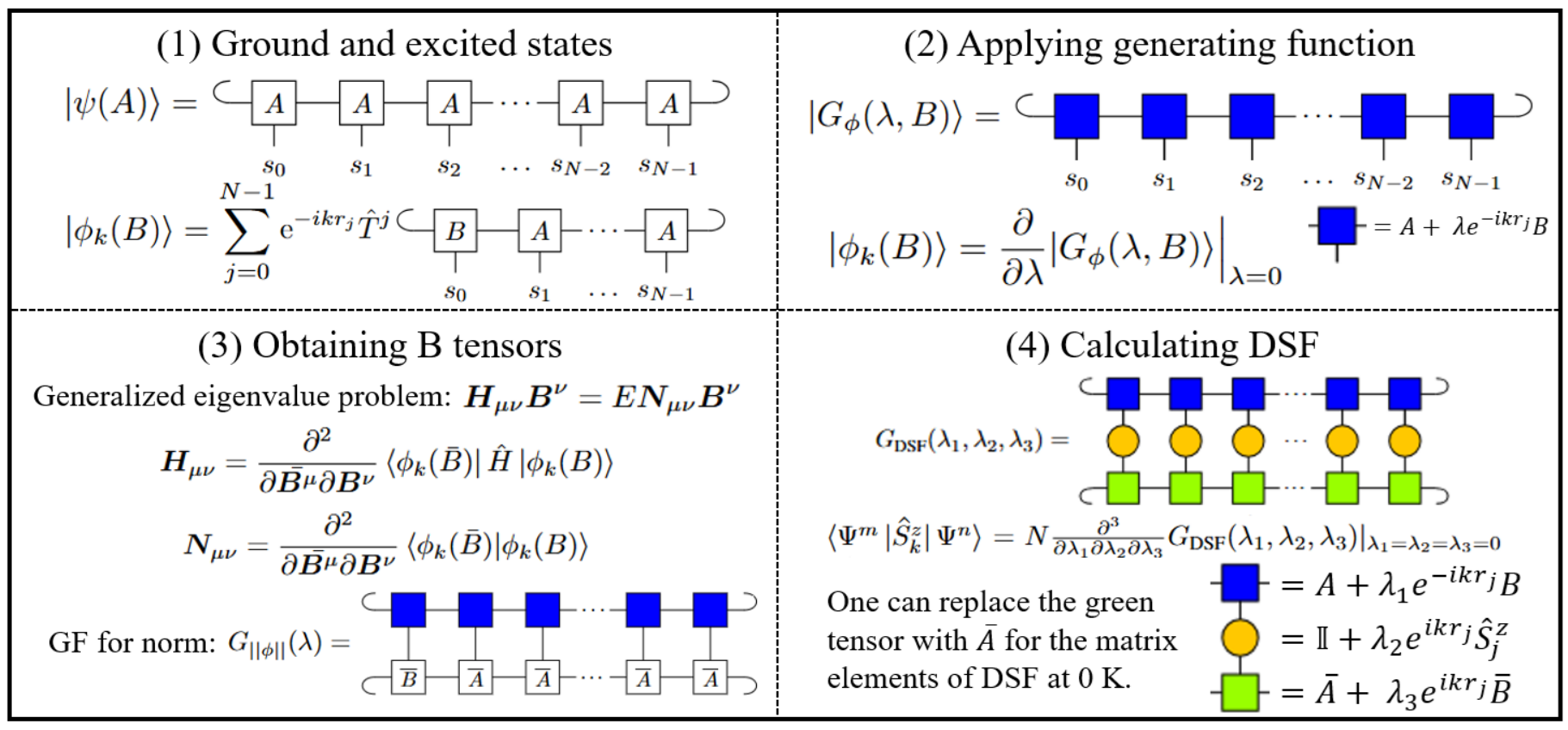}

  \caption{Overview of our algorithm. (1) The translationally invariant ground state is represented by an MPS with a single tensor $A$ forming a ring under periodic boundary conditions. Excited states are constructed as Bloch states with a wave number $k$ by inserting an impurity tensor $B$.
  (2) After optimizing $A$ via automatic differentiation~\cite{Liao2019}, excited states are generated by defining the GFMPS $|G_{\phi}(\lambda, B)\rangle$ and taking its derivative with respect to $\lambda$.
  (3) The tensor $B$ is obtained from a generalized eigenvalue problem, where the effective Hamiltonian $\bm{H}_{\mu\nu}$ and effective norm $\bm{N}_{\mu\nu}$ follow from differentiating with respect to $B$ and its complex conjugate, $\bar{B}$. The indices $\mu$ and $\nu$ label the components contracted with $B$ and $\bar{B}$, respectively, and their dimensionality matches the size of the Hilbert space spanned by the accessible quantum states. This procedure yields an orthogonal set of $B$ tensors, and translational invariance ensures that the GFMPS needs to be evaluated only in the ket layer.
  (4) Matrix elements, $\left\langle \Psi^{m} \left| \hat{S}^{z}_k \right| \Psi^n \right\rangle$, require three derivatives of $G_{\text{DSF}}(\lambda_1,\lambda_2, \lambda_3)$. At zero temperature, one fewer derivative is needed, as shown in Eq.~(\ref{eq:DSF_0K}). Applying Lorentzian broadening gives the final DSF results.
  }
  \label{fig:method}
\end{figure*}

In this work, we introduce a new algorithm for numerically evaluating the DSF using the matrix product state (MPS) framework. 
The approach requires expressing the Gibbs state in an MPS representation and subsequently performing the corresponding thermal averages.
For evaluating finite-temperature properties, we adopt the Gibbs state shown in Eq.~(\ref{Gibbs_state}), and the expectation value of operator $\hat{O}$ can then be computed by tr$(\rho_\text{G}\hat{O})$.
After solving for the eigenstates of our target Hamiltonian, we then construct the Gibbs state as
\begin{equation}
\rho_\text{G}=\frac{1}{Z}\sum_n e^{-\beta \epsilon_n}|\Psi^n\rangle\langle\Psi^n|, 
\label{Gibbs_state2}
\end{equation}
where $\epsilon_n=E_n-E_0$, the energy difference of $|\Psi^n\rangle$ to the ground state~($n$ represents the $n^{\text{th}}$ eigenstate), and $Z=\sum_n e^{-\beta \epsilon_n}$ is the partition function.
Accordingly, the DSF is evaluated according to the following equation 
\begin{equation}
\scalebox{1.0}{$
    S^{z}(\textbf{k}, \omega; \beta) = \frac{\sum_{n} e^{-\beta\epsilon_n} \sum_{m} \left| \left\langle \Psi^m \left| \hat{S}^{z}_\textbf{k} \right| \Psi^n \right\rangle \right|^2}{Z}\delta\left( \omega - E_m + E_n \right),
$}
\label{eq:finitetempsqw}
\end{equation}
where $\beta$ denotes the inverse temperature and $E_m$ is the $m^{\text{th}}$ eigenvalue. $\hat{S}_\textbf{k}^z$ denotes the Fourier transformed spin operator in $z$-axis.

For obtaining DSF, the most important ingredient lies on preparation of eigenstates and eigenvalues.
As proposed in our previous work we have developed an eigen-solver for one-dimensional~(1D) system with periodic boundary condition~\cite{Tu2021}.
We will directly adopt that package for collecting many eigenstates.
However, constructing the excited states, $\ket{\phi_k(B)}$, requires summing over all sites where $A$ tensor is replaced by $B$.
Here, the $A$ tensor is used to represent the ground state with translational invariance, while the $B$ tensor is the impurity tensor representing the local excitation (see Eqs.~(\ref{eq:uniformMPS}) and (\ref{eq:excitation})).
This summation leads to an extra order in computational cost, making the calculation potentially intractable as the size increases.
Therefore, we adopt the GFMPS that replaces the action of summation by differentiation~\cite{Tu2021, Tu2024}.
We first define the GFMPS for $\ket{\phi_k(B)}$
\begin{equation}
|G_{\phi}(\lambda, B)\rangle =
\begin{diagram}
\draw (0.5, 1.5) .. controls (0, 1.5) and (0, 2) .. (0.5, 2);
\draw (0.5, 1.5) -- (1, 1.5); \filldraw[fill={rgb, 255: red, 0; green, 0; blue, 255}] (1, 2) rectangle (2, 1);\draw (1.5, 1.3) node (X) {};
\draw (2, 1.5) -- (3, 1.5); \filldraw[fill={rgb, 255: red, 0; green, 0; blue, 255}] (3, 2) rectangle (4, 1);
\draw (4, 1.5) -- (5, 1.5); \filldraw[fill={rgb, 255: red, 0; green, 0; blue, 255}] (5, 2) rectangle (6, 1); 
\draw (6, 1.5) -- (6.5, 1.5); 
\draw (7, 1.5) node {$\ldots$};
\draw (7.5, 1.5) -- (8, 1.5); \filldraw[fill={rgb, 255: red, 0; green, 0; blue, 255}] (8, 2) rectangle (9, 1);
\draw (9, 1.5) -- (10, 1.5); \filldraw[fill={rgb, 255: red, 0; green, 0; blue, 255}] (10, 2) rectangle (11, 1);
\draw (11, 1.5) -- (11.5, 1.5);
\draw (11.5, 1.5) .. controls (12, 1.5) and (12, 2) .. (11.5, 2);
\draw (1.5, 1) -- (1.5, 0.5); \draw (3.5, 1) -- (3.5, 0.5); \draw (5.5, 1) -- (5.5, 0.5);
\draw (8.5, 1) -- (8.5, 0.5); \draw (10.5, 1) -- (10.5, 0.5);
\draw (1.5, 1) -- (1.5, 0.5); \draw (1.5, 0) node {$s_0$};
\draw (3.5, 1) -- (3.5, 0.5); \draw (3.5, 0) node {$s_1$};
\draw (5.5, 1) -- (5.5, 0.5); \draw (5.5, 0) node {$s_2$};
\draw (7, 0) node {$\ldots$};
\draw (8.5, 1) -- (8.5, 0.5); \draw (8.5, 0) node {$s_{N-2}$};
\draw (10.5, 1) -- (10.5, 0.5); \draw (10.5, 0) node {$s_{N-1}$};
\end{diagram},
\label{eq:excitationGen}
\end{equation}
where each blue tensor, $G_j(\lambda, B)$, is equal to $A+\lambda\mathrm{e}^{-ikr_j}B$ with $r_j$ indicating the position of the $j^{\text{th}}$ spin, $s_j$, and $\lambda\in\mathbb{R}$. $N$ indicates the system size.
It can be shown that 
\begin{equation}
|\phi_k(B)\rangle = \frac{\partial}{\partial\lambda}|G_{\phi}(\lambda, B)\rangle\Bigr |_{\lambda=0},
\label{eq:excitationDev}
\end{equation}
and the numerical differentiation can be well calculated by the automatic differentiation.

Due to the translational invariance, moreover, we can reduce the times of derivative by 1. For example, the norm of an excited state can be calculated through $||\phi||^2\equiv\langle\phi_{k}(B)|\phi_{k}(B)\rangle=\frac{\partial^2}{\partial\lambda'\partial\lambda}\langle G_{\phi}(\lambda', \bar{B})|G_{\phi}(\lambda, B)\rangle\Bigr |_{\lambda'=\lambda=0}$.
With the translational invariance, we can define the corresponding generating function~(GF) by 
\begin{equation}
G_{||\phi||}(\lambda)=
\begin{diagram}
\draw (0.5, 1.5) .. controls (0, 1.5) and (0, 2) .. (0.5, 2);
\draw (0.5, 1.5) -- (1, 1.5); \filldraw[fill={rgb, 255: red, 0; green, 0; blue, 255}] (1, 2) rectangle (2, 1);
\draw (2, 1.5) -- (3, 1.5); \filldraw[fill={rgb, 255: red, 0; green, 0; blue, 255}] (3, 2) rectangle (4, 1);
\draw (4, 1.5) -- (5, 1.5); \filldraw[fill={rgb, 255: red, 0; green, 0; blue, 255}] (5, 2) rectangle (6, 1);
\draw (6, 1.5) -- (6.5, 1.5); 
\draw (7, 1.5) node {$\ldots$};
\draw (7.5, 1.5) -- (8, 1.5); \filldraw[fill={rgb, 255: red, 0; green, 0; blue, 255}] (8, 2) rectangle (9, 1);
\draw (9, 1.5) -- (10, 1.5); \filldraw[fill={rgb, 255: red, 0; green, 0; blue, 255}] (10, 2) rectangle (11, 1);
\draw (11, 1.5) -- (11.5, 1.5);
\draw (11.5, 1.5) .. controls (12, 1.5) and (12, 2) .. (11.5, 2);
\draw (1.5, 1) -- (1.5, 0); \draw (3.5, 1) -- (3.5, 0); \draw (5.5, 1) -- (5.5, 0);
\draw (8.5, 1) -- (8.5, 0); \draw (10.5, 1) -- (10.5, 0);
\draw (1.5, 0.3) node (X) {};
\draw (0.5, -0.5) .. controls (0, -0.5) and (0, -1) .. (0.5, -1);
\draw (0.5, -0.5) -- (1, -0.5); \draw[] (1, 0) rectangle (2, -1); \draw (1.5, -0.5) node {$\overline{B}$};
\draw (2, -0.5) -- (3, -0.5); \draw[] (3, 0) rectangle (4, -1); \draw (3.5, -0.5) node {$\overline{A}$};
\draw (4, -0.5) -- (5, -0.5); \draw[] (5, 0) rectangle (6, -1); \draw (5.5, -0.5) node {$\overline{A}$}; \draw (6, -0.5) -- (6.5, -0.5); 
\draw (7, -0.5) node {\ldots};
\draw (7.5, -0.5) -- (8, -0.5); \draw[] (8, 0) rectangle (9, -1); \draw (8.5, -0.5) node {$\overline{A}$};
\draw (9, -0.5) -- (10, -0.5); \draw[] (10, 0) rectangle (11, -1); \draw (10.5, -0.5) node {$\overline{A}$};
\draw (11, -0.5) -- (11.5, -0.5);
\draw (11.5, -0.5) .. controls (12, -0.5) and (12, -1) .. (11.5, -1);
\end{diagram},
\label{eq:normGen}
\end{equation}
and the norm can be obtained as $||\phi||^2=N\frac{\partial}{\partial\lambda}G_{||\phi||}(\lambda)\Bigr |_{\lambda=0}$. From this point onward, the dependence of the GFs on $B$ is omitted, as no further differentiation with respect to $B$ is required once the generalized eigenvalue problem has been solved.

In fact, GF can not only be helpful during the preparation of the states, but it also reduces the computational complexity when calculating the physical observables.
For obtaining the DSF, one needs to consider the off-diagonal values for the Fourier-transformed $\hat{S}^z$ operator, defined as $\hat{S}^{z}_k = \frac{1}{\sqrt{N}}\sum_{j} e^{ikr_j}\hat{S}^{z}_j$.
The corresponding GF for such an operator can be dipicted as
\begin{equation}
\hat{G}_{S}(\lambda)=
\begin{diagram}
\draw (1.5, 1) -- (1.5, 0.5); \filldraw[fill={rgb, 255: red, 255; green, 200; blue, 0}] (1.5, 0) circle (0.5); \draw (1.5, -0.5) -- (1.5,-1);
\draw (3.5, 1) -- (3.5, 0.5); \filldraw[fill={rgb, 255: red, 255; green, 200; blue, 0}] (3.5, 0) circle (0.5); \draw (3.5, -0.5) -- (3.5, -1);
\draw (5.5, 1) -- (5.5, 0.5); \filldraw[fill={rgb, 255: red, 255; green, 200; blue, 0}] (5.5, 0) circle (0.5); \draw (5.5, -0.5) -- (5.5, -1);
\draw (8.5, 1) -- (8.5, 0.5); \filldraw[fill={rgb, 255: red, 255; green, 200; blue, 0}] (8.5, 0) circle (0.5); \draw (8.5, -0.5) -- (8.5, -1); 
\draw (10.5, 1) -- (10.5, 0.5); \filldraw[fill={rgb, 255: red, 255; green, 200; blue, 0}] (10.5, 0) circle (0.5); \draw (10.5, -0.5) -- (10.5, -1);
\draw (1.5, -0.2) node (X) {};
\draw (7, 0) node {$\ldots$};
\draw (1.5, 1.5) node {$s_0$};
\draw (3.5, 1.5) node {$s_1$};
\draw (5.5, 1.5) node {$s_2$};
\draw (8.5, 1.5) node {$s_{N-2}$};
\draw (10.5, 1.5) node {$s_{N-1}$};
\draw (7, 1.5) node {$\ldots$};
\end{diagram},
\label{eq:sqwGen}
\end{equation}
where the orange operators are defined by $\hat{G}^{z}_j(\lambda)=\mathbb{I}+\lambda\mathrm{e}^{ikr_j}\hat{S}^{z}_j$. 
$\hat{S}^{z}_k$ can then be obtained through a derivative to $\lambda$, $\hat{S}^{z}_k = \frac{1}{\sqrt{N}}\frac{d}{d\lambda}\hat{G}_{S}(\lambda)\Bigr |_{\lambda=0}$.

In order to calculate DSF, multiple derivatives are then required.
For example, DSF at 0 K is given by $S^{z}(k, \omega) = \sum_{n} \left| M^{z}_{k} \right|^{2} \delta(\omega - E^{k}_{n} + E_{0})$, where $
M^{z}_{k} = \langle \phi_k (B_{n}) | \hat{S}^{z}_{k} | \psi(A) \rangle$, and $E_0$ and $E_n^k$ denote the ground- and excited-state energies, respectively.
Note that here the extra upper index $k$ for $E_n$ indicates that such an eigenvalue is obtained with fixed momentum $k$.
The corresponding GF for the non-normalized DSF is
\begin{equation}
G_{\text{DSF}}^0(\lambda_1,\lambda_2)=
\begin{diagram}
\draw (0.5, 1.5) .. controls (0, 1.5) and (0, 2) .. (0.5, 2);
\draw (0.5, 1.5) -- (1, 1.5); \filldraw[fill={rgb, 255: red, 0; green, 0; blue, 255}] (1, 2) rectangle (2, 1);
\draw (2, 1.5) -- (3, 1.5); \filldraw[fill={rgb, 255: red, 0; green, 0; blue, 255}] (3, 2) rectangle (4, 1);
\draw (4, 1.5) -- (5, 1.5); \filldraw[fill={rgb, 255: red, 0; green, 0; blue, 255}] (5, 2) rectangle (6, 1);
\draw (6, 1.5) -- (6.5, 1.5); 
\draw (7, 1.5) node {$\ldots$};
\draw (7.5, 1.5) -- (8, 1.5); \filldraw[fill={rgb, 255: red, 0; green, 0; blue, 255}] (8, 2) rectangle (9, 1);
\draw (9, 1.5) -- (10, 1.5); \filldraw[fill={rgb, 255: red, 0; green, 0; blue, 255}] (10, 2) rectangle (11, 1);
\draw (11, 1.5) -- (11.5, 1.5);
\draw (11.5, 1.5) .. controls (12, 1.5) and (12, 2) .. (11.5, 2);
\draw (1.5, 1) -- (1.5, 0); \draw (3.5, 1) -- (3.5, 0); \draw (5.5, 1) -- (5.5, 0);
\draw (8.5, 1) -- (8.5, 0); \draw (10.5, 1) -- (10.5, 0);
\draw (1.5, 0.3) node (X) {};

\draw (1.5, 1) -- (1.5, 0.5); \filldraw[fill={rgb, 255: red, 255; green, 200; blue, 0}] (1.5, 0) circle (0.5); \draw (1.5, -0.5) -- (1.5,-1);
\draw (3.5, 1) -- (3.5, 0.5); \filldraw[fill={rgb, 255: red, 255; green, 200; blue, 0}] (3.5, 0) circle (0.5); \draw (3.5, -0.5) -- (3.5, -1);
\draw (5.5, 1) -- (5.5, 0.5); \filldraw[fill={rgb, 255: red, 255; green, 200; blue, 0}] (5.5, 0) circle (0.5); \draw (5.5, -0.5) -- (5.5, -1);
\draw (8.5, 1) -- (8.5, 0.5); \filldraw[fill={rgb, 255: red, 255; green, 200; blue, 0}] (8.5, 0) circle (0.5); \draw (8.5, -0.5) -- (8.5, -1); 
\draw (10.5, 1) -- (10.5, 0.5); \filldraw[fill={rgb, 255: red, 255; green, 200; blue, 0}] (10.5, 0) circle (0.5); \draw (10.5, -0.5) -- (10.5, -1);
\draw (1.5, -0.2) node (X) {};
\draw (7, 0) node {$\ldots$};
\draw (1.5, 1.5);
\draw (3.5, 1.5);
\draw (5.5, 1.5);
\draw (8.5, 1.5);
\draw (10.5, 1.5); 
\draw (7, 1.5) node {$\ldots$};

\draw (0.5, -1.5) .. controls (0, -1.5) and (0, -2) .. (0.5, -2);
\draw (0.5, -1.5) -- (1, -1.5); \draw[] (1, -1) rectangle (2, -2); \draw (1.5, -1.5) node {$\overline{A}$};
\draw (2, -1.5) -- (3, -1.5); \draw[] (3, -1) rectangle (4, -2); \draw (3.5, -1.5) node {$\overline{A}$};
\draw (4, -1.5) -- (5, -1.5); \draw[] (5, -1) rectangle (6, -2); \draw (5.5, -1.5) node {$\overline{A}$}; \draw (6, -1.5) -- (6.5, -1.5); 
\draw (7, -1.5) node {\ldots};
\draw (7.5, -1.5) -- (8, -1.5); \draw[] (8, -1) rectangle (9, -2); \draw (8.5, -1.5) node {$\overline{A}$};
\draw (9, -1.5) -- (10, -1.5); \draw[] (10, -1) rectangle (11, -2); \draw (10.5, -1.5) node {$\overline{A}$};
\draw (11, -1.5) -- (11.5, -1.5);
\draw (11.5, -1.5) .. controls (12, -1.5) and (12, -2) .. (11.5, -2);
\end{diagram},
\label{eq:DSF_0K}
\end{equation}
and $M^{z}_{k} = N \frac{\partial^2}{\partial \lambda_1 \partial \lambda_2}G_{\text{DSF}}^0(\lambda_1,\lambda_2) |_{\lambda_1=\lambda_2=0}$.

As for the finite-temperature DSF, on the other hand, we follow Eq.~(\ref{eq:finitetempsqw}) for its calculation.
In this formulation, we then need a three-time derivative to obtain the off-diagonal matrix elements, $\left\langle \Psi^{m} \left| \hat{S}^{z}_k \right| \Psi^n \right\rangle$, which is equal to $N \frac{\partial^3}{\partial \lambda_1 \partial \lambda_2 \partial \lambda_3}G_{\text{DSF}}(\lambda_1,\lambda_2,\lambda_3) |_{\lambda_1=\lambda_2=\lambda_3=0}$ and
\begin{equation}
G_{\text{DSF}}(\lambda_1,\lambda_2, \lambda_3)=
\begin{diagram}
\draw (0.5, 1.5) .. controls (0, 1.5) and (0, 2) .. (0.5, 2);
\draw (0.5, 1.5) -- (1, 1.5); \filldraw[fill={rgb, 255: red, 0; green, 0; blue, 255}] (1, 2) rectangle (2, 1);
\draw (2, 1.5) -- (3, 1.5); \filldraw[fill={rgb, 255: red, 0; green, 0; blue, 255}] (3, 2) rectangle (4, 1);
\draw (4, 1.5) -- (5, 1.5); \filldraw[fill={rgb, 255: red, 0; green, 0; blue, 255}] (5, 2) rectangle (6, 1);
\draw (6, 1.5) -- (6.5, 1.5); 
\draw (7, 1.5) node {$\ldots$};
\draw (7.5, 1.5) -- (8, 1.5); \filldraw[fill={rgb, 255: red, 0; green, 0; blue, 255}] (8, 2) rectangle (9, 1);
\draw (9, 1.5) -- (10, 1.5); \filldraw[fill={rgb, 255: red, 0; green, 0; blue, 255}] (10, 2) rectangle (11, 1);
\draw (11, 1.5) -- (11.5, 1.5);
\draw (11.5, 1.5) .. controls (12, 1.5) and (12, 2) .. (11.5, 2);
\draw (1.5, 1) -- (1.5, 0); \draw (3.5, 1) -- (3.5, 0); \draw (5.5, 1) -- (5.5, 0);
\draw (8.5, 1) -- (8.5, 0); \draw (10.5, 1) -- (10.5, 0);
\draw (1.5, 0.3) node (X) {};

\draw (1.5, 1) -- (1.5, 0.5); \filldraw[fill={rgb, 255: red, 255; green, 200; blue, 0}] (1.5, 0) circle (0.5); \draw (1.5, -0.5) -- (1.5,-1);
\draw (3.5, 1) -- (3.5, 0.5); \filldraw[fill={rgb, 255: red, 255; green, 200; blue, 0}] (3.5, 0) circle (0.5); \draw (3.5, -0.5) -- (3.5, -1);
\draw (5.5, 1) -- (5.5, 0.5); \filldraw[fill={rgb, 255: red, 255; green, 200; blue, 0}] (5.5, 0) circle (0.5); \draw (5.5, -0.5) -- (5.5, -1);
\draw (8.5, 1) -- (8.5, 0.5); \filldraw[fill={rgb, 255: red, 255; green, 200; blue, 0}] (8.5, 0) circle (0.5); \draw (8.5, -0.5) -- (8.5, -1); 
\draw (10.5, 1) -- (10.5, 0.5); \filldraw[fill={rgb, 255: red, 255; green, 200; blue, 0}] (10.5, 0) circle (0.5); \draw (10.5, -0.5) -- (10.5, -1);
\draw (1.5, -0.2) node (X) {};
\draw (7, 0) node {$\ldots$};
\draw (1.5, 1.5);
\draw (3.5, 1.5);
\draw (5.5, 1.5);
\draw (8.5, 1.5);
\draw (10.5, 1.5); 
\draw (7, 1.5) node {$\ldots$};

\draw (0.5, -1.5) .. controls (0, -1.5) and (0, -2) .. (0.5, -2);
\draw (0.5, -1.5) -- (1, -1.5);

\draw[fill=green!40!yellow, draw=black] (1, -1) rectangle (2, -2);

\draw (2, -1.5) -- (3, -1.5);
\draw[fill=green!40!yellow, draw=black] (3, -1) rectangle (4, -2);

\draw (4, -1.5) -- (5, -1.5);
\draw[fill=green!40!yellow, draw=black] (5, -1) rectangle (6, -2);

\draw (6, -1.5) -- (6.5, -1.5);
\draw (7, -1.5) node {\ldots};
\draw (7.5, -1.5) -- (8, -1.5);

\draw[fill=green!40!yellow, draw=black] (8, -1) rectangle (9, -2);

\draw (9, -1.5) -- (10, -1.5);
\draw[fill=green!40!yellow, draw=black] (10, -1) rectangle (11, -2);

\draw (11, -1.5) -- (11.5, -1.5);
\draw (11.5, -1.5) .. controls (12, -1.5) and (12, -2) .. (11.5, -2);

\end{diagram},
\label{eq:DSF_finite}
\end{equation}
where for the green tensors we have $G_j(\lambda_3, \bar{B}) = \bar{A}+\lambda_3\mathrm{e}^{-ikr_j}\bar{B}$.
Note that here $|\Psi^{m \vee n}\rangle$ can be either ground or excited states.
It is thus clear that compared to the DSF at 0 K now the required computational complexity becomes much larger not only due to one extra time for the differentiation, but also because of the fact that we need to take into account numerous excited states in the bra layer.
Fortunately, the matrix elements do not alter along with the temperature and thus we can store the information for DSF at various temperatures.
By taking the finite-temperature average with Lorentzian broadening, at last we obtain DSF at different temperatures for further comparison with the exact diagonalization~(ED) or experimental results.

The overall procedure is illustrated in Fig.~\ref{fig:method}, where we demonstrate how one can generate the excited states from ground state formulated through the GFMPS approach.
Within this framework, the construction of a new tensor $B$ is required, which is obtained by solving a generalized eigenvalue problem. 
The resulting $B$ tensors are then employed to build the excited states, from which the DSF is evaluated by combining the excited states with the operator relevant to the DSF calculation. 
In our algorithm, evaluating a single matrix element incurs a computational cost of $\mathcal{O}(Nd\chi^5)$, where $\chi$ is the bond dimension and $d$ is the local Hilbert‑space dimension.
For finite‑temperature DSF calculations, however, contributions from numerous excited states must be included, increasing the overall cost to $\mathcal{O}(M^2Nd\chi^5)$, where $M$ denotes the number of retained states.
Although the computation becomes increasingly demanding as $M$ grows, only matrix elements that satisfy momentum conservation can yield nonzero contributions.
Specifically, let the bra state carry momentum $k_1$, the ket state momentum $k_2$, and the Fourier‑transformed operator $\hat{S}^{z}_k$ carry momentum $k_3$.
Then the Momentum conservation requires $k_1=k_2+k_3$. Matrix elements that do not satisfy this condition vanish identically. By exploiting this constraint, we can systematically discard zero matrix elements and thereby reduce the effective computational cost.

Additional technical details, including the procedure for constructing the impurity tensors, are provided in Section~\ref{sec:method}. 
For those who might be interested, we also provide the TN library for generating the 1D finite-temperature Gibbs states using MPS~\footnote{See \href{https://github.com/tomoya1017/GFMPS}{https://github.com/tomoya1017/GFMPS} for code implementation}.
We now turn to the presentation of our computational results in the subsequent sections.

\begin{figure*}
  \centering
  \includegraphics[scale=0.55]{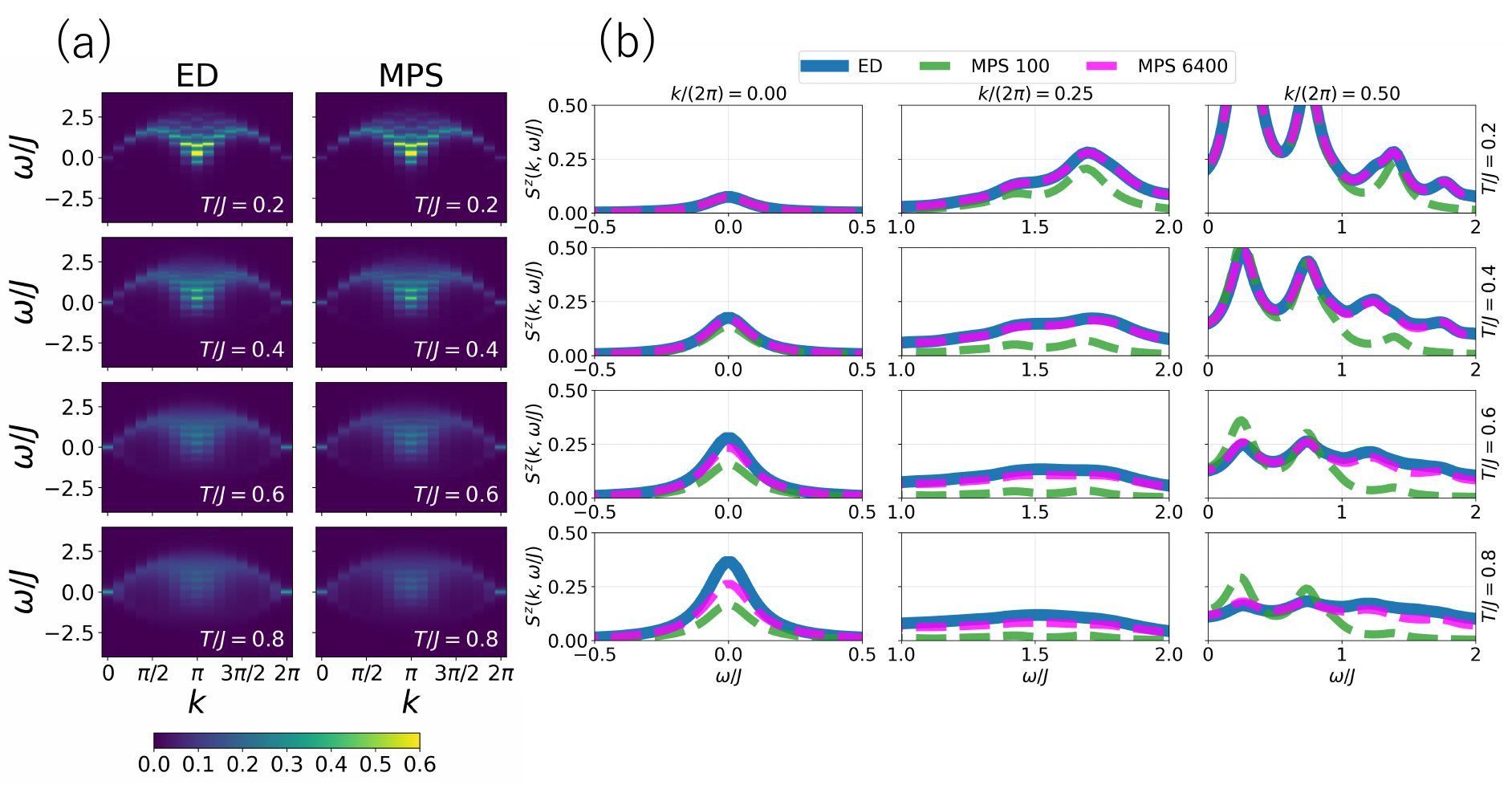}
  \caption{The DSF of the Heisenberg chain with $S=1/2$ and $N=16$, computed using ED and GFMPS methods are shown. Panel (a) presents color contour maps of the DSF at temperatures $T/J=0.2$, 0.4, 0.6, and 0.8~(from top to bottom). Panel (b) displays DSF cuts at fixed wave numbers $k/2\pi=0$, 0.25, and 0.5~(from left to right) for the same set of temperatures. To illustrate the influence of the number of retained states in the MPS calculations, we compare results obtained using 100 states (green dashed lines) and 6400 states (pink dashed lines).
  }
  \label{fig:spin_half_qcut}
\end{figure*}

\begin{figure*}
  \centering
  \includegraphics[scale=0.55]{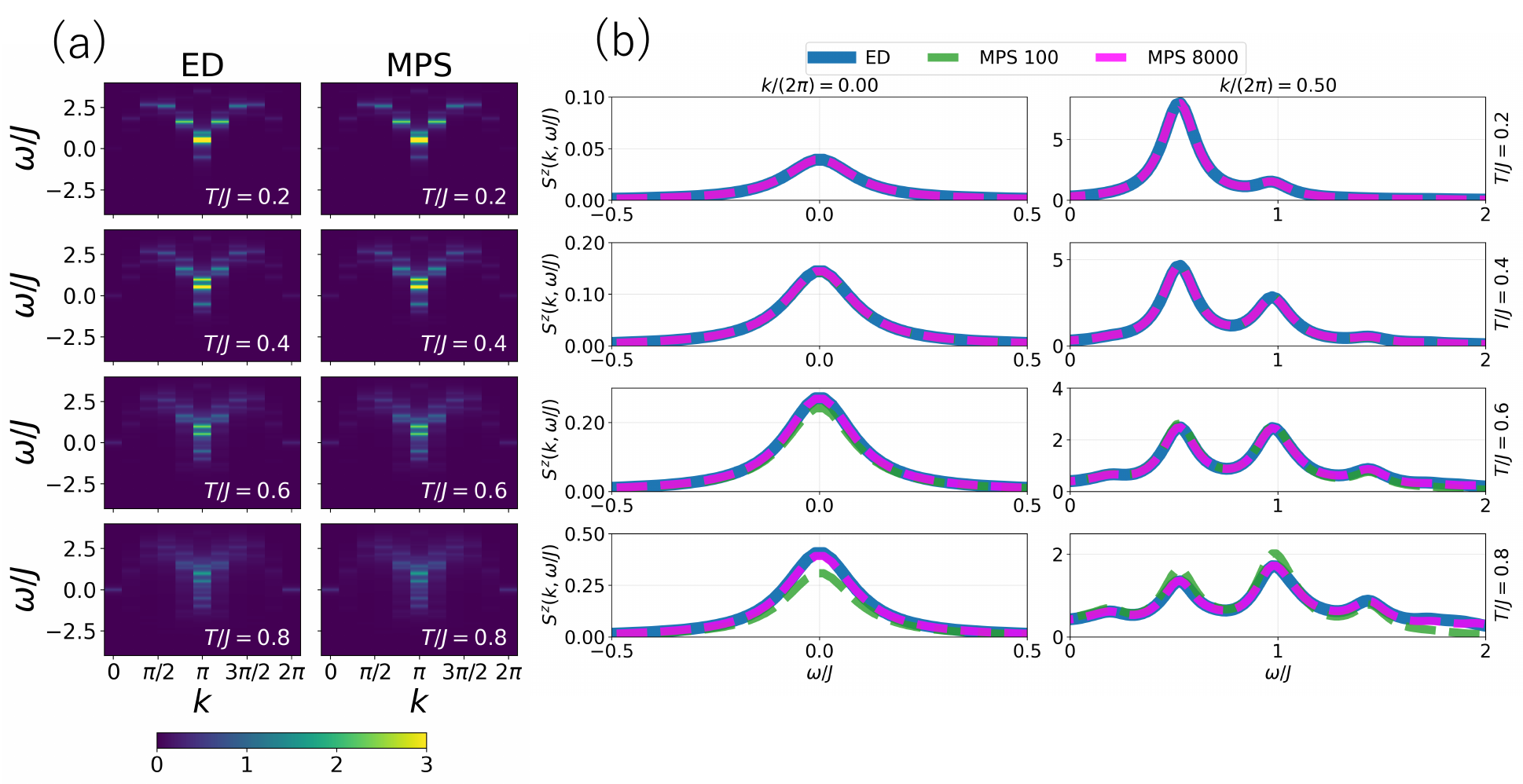}
  \caption{Similar figures as Fig.~\ref{fig:spin_half_qcut} for the $S=1$ Heisenberg model with $N=10$. Panel (a) now presents color contour maps of DSF at temperatures $T/J=0.2$, 0.4, 0.6, and 0.8~(from top to bottom). Panel (b) displays DSF cuts at fixed wave numbers $k/2\pi=0$ and 0.5~(from left to right) for the set of temperatures in (a). Here we also compare results obtained using 100 states (green dashed lines) and 8000 states (pink dashed lines).}
  \label{fig:spin_one_qcut}
\end{figure*}

\subsection{Benchmark results}
\label{subsec:finite_temp}

In order to validate the effectiveness of our method, we first conduct calculations in smaller systems to benchmark with those by ED.
We consider the 1D Heisenberg model whose Hamiltonian is written as
\begin{equation}
    H = J \sum_{i=0}^{N-1}(\hat{S}_i^x\hat{S}_{i+1}^x + \hat{S}_i^y\hat{S}_{i+1}^y +\hat{S}_i^z\hat{S}_{i+1}^z),
\label{eq:hamiltonian}
\end{equation}
where $J$ is the antiferromagnetic exchange interaction between spin operators and throughout this paper we set it to be the energy unit. Here we adopt the periodic boundary condition so that $\hat{\textbf{S}}_{N}=\hat{\textbf{S}}_0$.
Specifically, we considered system size $N=16$ for $S=1/2$ and $N=10$ for $S=1$.
In the following data we have replaced the delta function in Eq.~(\ref{eq:finitetempsqw}) by a normalized Lorentzian with broadening width $\sigma=0.1J$.

In Figs.~\ref{fig:spin_half_qcut} and \ref{fig:spin_one_qcut}, we present a comparison between the DSF obtained via ED and that computed using our method for various wave numbers $k$ and temperatures $T/J$ for $S=1/2$ and 1. 
Figs.~\ref{fig:spin_half_qcut} and \ref{fig:spin_one_qcut}(a) demonstrate the DSF calculated with the maximum number of states (6400 for $S=1/2$ and 8000 for $S=1$), plotted as a function of $\omega/J$ to momentum $k$.
Here and below the number of states refers to the number of eigenstates of the generalized eigenvalue problem, the Step 3 in Fig.~\ref{fig:method}, ordered with increasing energy.
At low temperatures, the spin-wave excitation leads to a DSF characterized by a pronounced peak at $k=\pi$. 
As the temperature increases, however, the spin arrangement becomes progressively disordered, and the peak correspondingly shifts toward $k=0$.
It is clear that our approach successfully captures the above-mentioned feature.
In Fig.~\ref{fig:spin_half_qcut} and \ref{fig:spin_one_qcut}(b) we provide several cuts through the DSF plots and assess the influence of the number of retained states.
When only 100 states are adopted, the agreement with ED progressively deteriorated as the temperature increases. In contrast, employing the maximum number of states results in good agreement with ED even at elevated temperatures.
Note that in our framework, the maximum number of states that can be retained is constrained by $N\chi^2(d-1)$~($d=2$ for $S=1/2$ and $d=3$ for $S=1$).
Throughout this work we set $\chi=20$.
We can then conclude that achieving closer agreement with ED requires not only increasing the bond dimension to enhance the approximation accuracy, but also expanding the number of states that can be retained within the method.

\begin{table}[H]
    \centering
    \scalebox{1.3}{
    \begin{tabular}{c|c|c|c} \hline\hline
        $S$ & Fidelity & $\epsilon/J$ & $k$   \\ \hline

        1/2 & {1.000000} & {0.445749} & $\pi$  \\ \hline

        1/2 & {0.999929} & {1.551006} & $0$
\\ \hline

        \:\:\:1/2\:\:\: & \:\:\:{0.999976}\:\:\: & {1.668897} & \:\:\:$\pi$\:\:\:
\\ \hline
         1/2 & {0.998620} & \:\:\:{2.576509}\:\:\: & $0$
\\ \hline
         1/2 & {0.999345} & {2.588040} & $\pi$
\\ \hline\hline
         1 & {0.999969} & {2.248460} & $0$
\\ \hline
        1 & {0.999735} & {4.798602} & $0$
\\ \hline\hline

    \end{tabular}}
    \caption{Fidelity for the non-degenerate states for $S=1/2$~($N=16$) and $S=1$~($N=10$). All the numbers are rounded to the seventh decimal place.}
    \label{tab:fidelity}
\end{table}

\begin{figure*}
  \centering
  \includegraphics[scale=0.65]{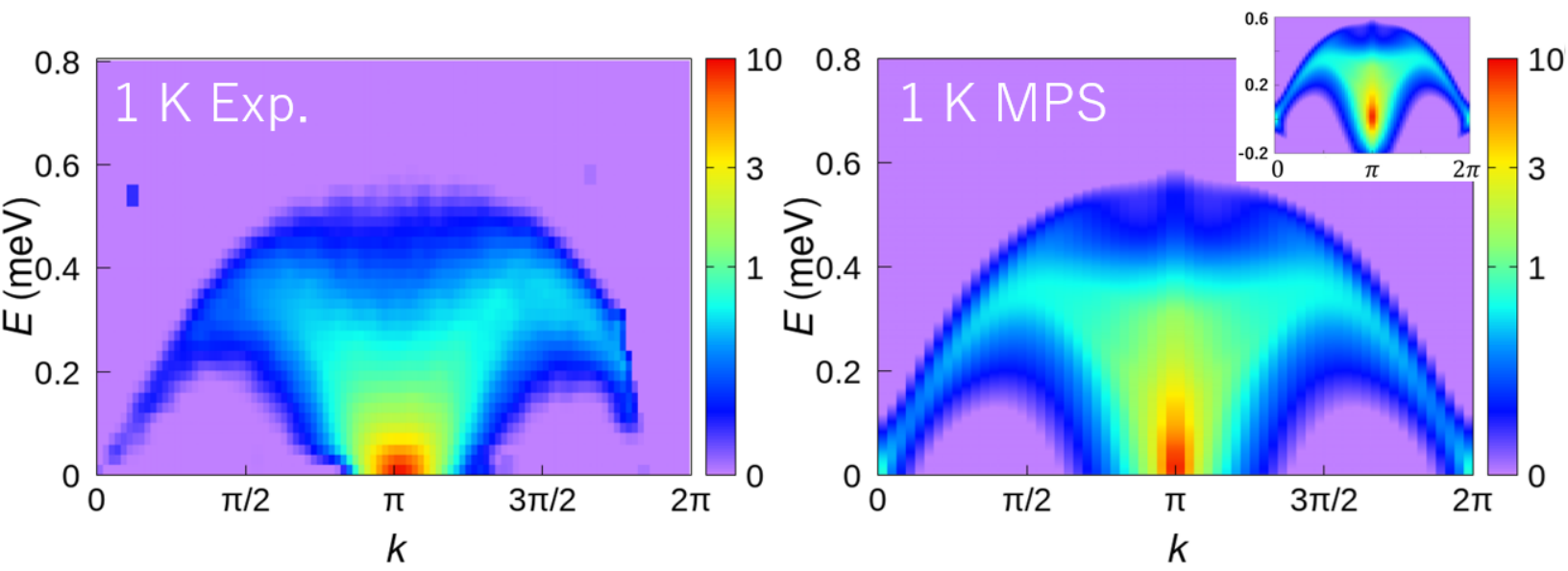}
  \caption{DSF obtained via experiment for YbAlO$_3$ at 1 K~\cite{L.S.Wu2019} (left), and that obtained form direct Gibbs-state construction using the GFMPS method for the $S=1/2$ Heisenberg chain with 64 sites (right). The relevant temperature scales can be estimated using the fact that the antiferromagnetic exchange coupling in YbAlO$_3$ is approximately $J\approx0.21$ meV~\cite{Radhakrishna1981}. Each dataset is first normalized by dividing by its own maximum. Afterward, the normalized values are scaled by a factor of 10, and the logarithm of the scaled data is taken to produce the contour plots. Our simulation retains $10^4$ states to ensure high‑accuracy results. We also demonstrate our DSF plot by MPS including negative $E$ in the upper panel on the right side within the MPS-derived DSF plot.
  }
  \label{fig:YbAlO3_dsf_temp}
\end{figure*}

As a more direct comparison, we compute the fidelity, a measure of how close an approximated state by MPS is to the corresponding exact quantum state, between our MPS eigenstates and ED eigenstates.
It can be computed as $|\langle\Psi_n^{\text{ED}}|\Psi_n^{\text{MPS}}\rangle|^2$, the squared inner product between the two $n^{\text{th}}$ eigenstates.
We sample the fidelity for low-energy non-degenerate states, shown in Table~\ref{tab:fidelity}.
It is clear that the fidelity remains to be very close to 1 within the energy range of interest, and it suggests that our method achieves very high approximation accuracy in the low-energy regime.
The fidelity for $S=1$ tends to be higher than that for $S=1/2$ and it is likely due to the fact that there is a finite energy gap even in the thermodynamic limit for $S=1$ Heisenberg model.

\subsection{$S=1/2$ Heisenberg spin chain with $N=64$}
\label{subsec:exp_benchmark}

While benchmarking with ED confirms that our method is reliable in the low-temperature regime, it is important to extend our simulation beyond the availability of ED and make some meaningful comparison with the experimental data.
It has been previously shown that a rare-earth perovskite, YbAlO$_3$, provides a good realization of $S=1/2$ Heisenberg chain with $J\approx0.21$ meV~\cite{Radhakrishna1981, L.S.Wu2019, Kish2025}.
This material has an orbital angular momentum $L=3$, spin angular momentum $S=1/2$, and total angular momentum $J=7/2$, giving an eightfold degeneracy.
Due to the crystal electric field, this degeneracy splits into four Kramers doublets~\cite{L.S.Wu2016, L.S.Wu2019}.
Since the ground-state doublet is well separated in energy from the first excited doublet, it is sufficient to consider only the ground-state doublet, which can be effectively modeled as an $S=1/2$ system.
Furthermore, previous studies have shown that there is a confinement-deconfinement transition at the Néel temperature~($T_\text{N}=0.88$ K), where the inter-chain coupling becomes unimportant and the material is well described by an antiferromagnetic Heisenberg model without any anisotropy~\cite{L.S.Wu2019}.

\begin{figure}
  \centering
  \includegraphics[scale=0.5]{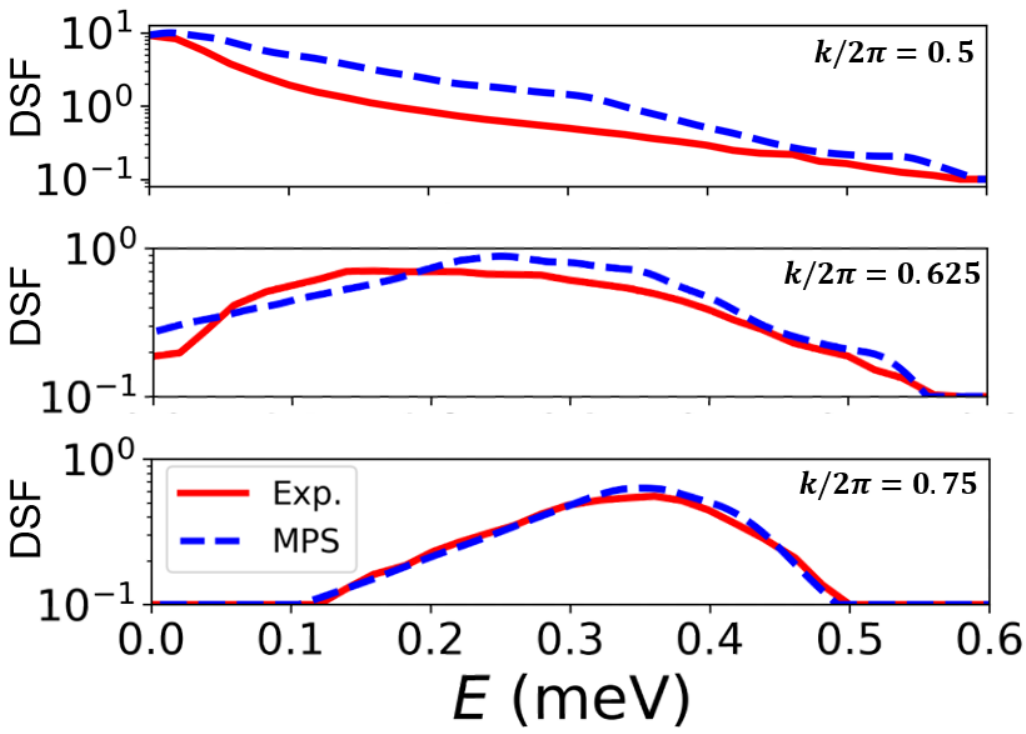}
  \caption{Several representative cuts through the DSF with fixed $k$ and $E \in[0,0.6]$ meV. The experimental data are shown in red color while MPS results in blue dashed lines.
  }
  \label{fig:YbAlO3_dsf_temp_qcut}
\end{figure}

\begin{figure*}
  \centering
 \includegraphics[scale=0.65]{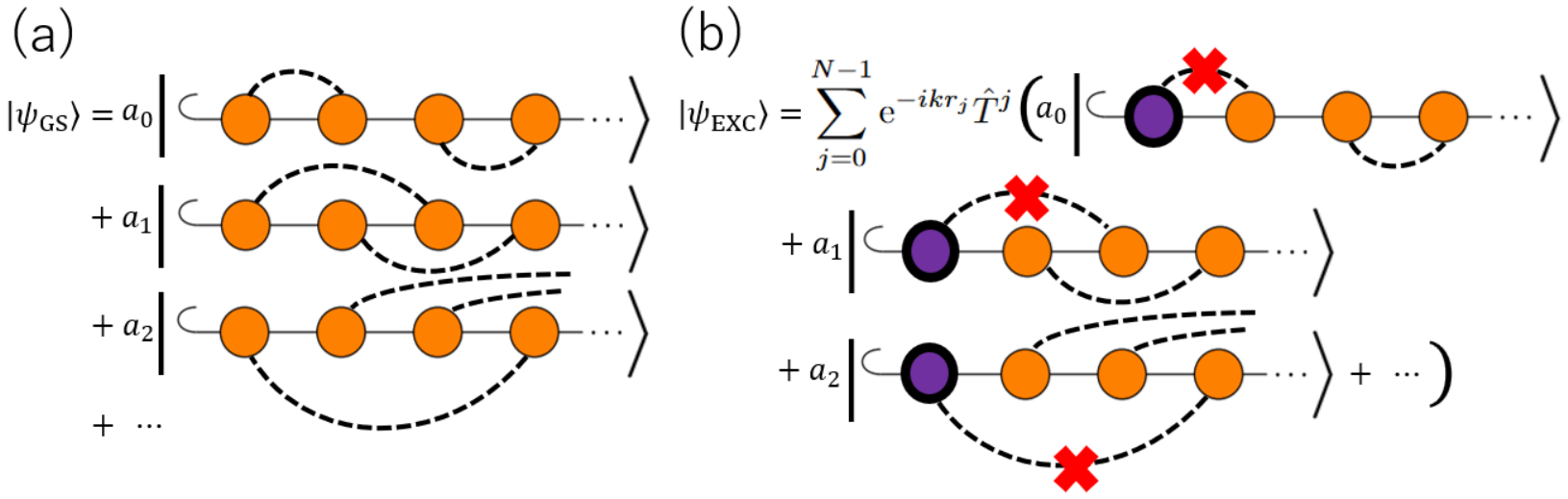}

  \caption{Schematic illustrations of the ground and excited states of the Heisenberg chain. (a) The ground state, described by a TLL, can be viewed as a quantum superposition of singlet‑pair configurations, indicated by black dashed lines. (b) Introducing a single‑mode excitation effectively replaces one site, breaking a singlet bond and liberating two unpaired spins. Owing to the superposed nature of the ground state, these unpaired spins naturally form the basis of the two‑spinon excitation picture. Furthermore, in the absence of interchain coupling, separating the two spinons does not incur additional energy, resulting in deconfined excitations.
  }
  \label{fig:spinon_exp}
\end{figure*}

In Fig.~\ref{fig:YbAlO3_dsf_temp}, we present our simulated DSF for the $N=64$, $S=1/2$ Heisenberg model with periodic boundary condition~(right), alongside the corresponding experimental data~(left). 
To ensure a meaningful comparison between the two datasets, we first normalize each set of raw data by its own maximum value. The normalized data are then uniformly scaled by a factor of 10, after which the logarithm is taken to generate the contour plots.
At finite temperatures, negative-frequency components acquire non-vanishing spectral weight~\cite{Kish2025}. This behavior is accurately reproduced in our numerical simulation, and the resulting spectrum including negative-$\omega$ is displayed in the inset located at the upper right of the MPS-derived DSF plot.
Note that such a calculation is beyond the reach of ED due to the exponential growth of the Hilbert‑space dimension.
While Fig.~\ref{fig:YbAlO3_dsf_temp} already shows that our simulated DSF reproduces the experimental spectrum with excellent accuracy, we provide a more detailed comparison in Fig.~\ref{fig:YbAlO3_dsf_temp_qcut}, where several representative cuts through the DSF are plotted. Across all these cuts, our results exhibit outstanding agreement with the experimental data.
It is worth noting that Ref.~\onlinecite{L.S.Wu2019} reported DMRG simulations performed at zero temperature, but their results show significant deviations from experiment.
For instance, the experimentally observed energy dispersion near $k=\pi/2$, $E=0.2$ meV is entirely absent in their DMRG spectrum~(see Fig.~3 of that reference).
This discrepancy is likely attributable to the absence of thermal effects in their calculation, while such effects can be accurately captured in our finite‑temperature simulation.

\subsection{Discussion}
\label{subsec:spinon}

In the above sections we have demonstrated that our algorithm combining the GFMPS and direct Gibbs state construction provides a good approximation for the low-temperature dynamical behavior for quantum spin chains.
It is important to notice that, however, in step (2) shown in Fig.~\ref{fig:method}, we only take one derivative to $\lambda$ for our GFMPS, which leads to the so-called single-mode excitation~\cite{Haegeman2013a}.
Conventionally, spinon excitation comes in pairs and manifests as two domain walls. It can be confined or deconfined depending on whether there is a energy gap in the system (see, e.g., explanation and demonstration in Ref.~\onlinecite{L.S.Wu2019}).
In principle, reproducing spinon excitation requires taking into account multiparticle excitation, which demands incorporating nonlocality into the ansatz.
Nevertheless, in this study we successfully reproduce the energy continuum as a signature of spinon excitations with only the single-mode excitation thanks to the property of the ground state.

It is well established that the overcomplete basis formed by all possible singlet‑pair (valence‑bond) configurations~\cite{Pauling1933} provides an accurate and efficient representation of the eigenstates and eigenvalues of 1D $S=1/2$ antiferromagnetic systems~\cite{Hulthén1938, Beach2006}. 
As illustrated in Fig.~\ref{fig:spinon_exp}(a), the ground state of the $S=1/2$ Heisenberg chain, known to be a Tomonaga–Luttinger liquid (TLL), can be naturally interpreted as a coherent superposition of these singlet coverings. 
In such a resonating-valence-bond–like state, the single‑mode excitation effectively breaks one of the singlets, producing two unpaired spins as shown in Fig.~\ref{fig:spinon_exp}(b).
Because this TLL contains singlets of all possible lengths, the two unpaired spins can separate arbitrarily far apart, giving rise to the characteristic picture of deconfined spinons.

At finite temperatures, however, multiparticle excitations become increasingly important and contribute substantially to the growth of entanglement. 
These contributions are not fully captured within our current algorithm, which is built around the simplified picture of deconfined spinons. 
This limitation likely explains why our results deviate more noticeably from the ED ones in the gapless $S=1/2$ Heisenberg model. 
While this indicates that an extension of the method may be required to accurately treat higher‑temperature regimes in critical systems, the present algorithm remains reliable at extremely low but finite temperatures.

\subsection{Conclusion remarks}
\label{subsec:Conclusion}

In this work, we propose a direct and efficient method for constructing the Gibbs state using GFMPS, bypassing the need for imaginary time evolution or purification. 
Our approach leverages a high-performance excited-state eigensolver developed in our previous work~\cite{Tu2021}, which enables the accurate computation of a large number of excited eigenstates across a broad energy spectrum. 
By systematically incorporating these eigenstates into the thermal ensemble, we construct the Gibbs-state ansatz and benchmark the reconstructed thermal observables against ED results. 
We find remarkable agreement when a sufficient number of excited states are included, particularly in the low-temperature regime where the contribution from low-lying states dominates. 
After enlarging the size of system beyond the attainability of ED, furthermore, we compared our results with available experimental data and observed strong consistency, underscoring the practical relevance and predictive power of our method.

This framework opens promising new avenues for investigating thermal properties of quantum systems with high precision, particularly in regimes where conventional finite-temperature tensor network methods encounter significant limitations. 
Techniques such as purification and METTS have been widely adopted for simulating thermal states, yet each comes with inherent trade-offs. As discussed in Ref.~\onlinecite{Binder2015}, purification generally offers superior performance in terms of accuracy and convergence, especially at moderate temperatures. 
However, METTS becomes increasingly efficient when the system exhibits a large energy gap relative to the temperature scale, making it particularly well-suited for probing low-temperature observables in gapped systems.
Despite this advantage, METTS suffers from severe entanglement growth during imaginary time evolution, especially in gapless or critical systems, which can lead to substantial computational overhead and reduced accuracy. 
In contrast, our proposed approach circumvents these challenges by leveraging explicit eigenstate information. When calculating the response functions, especially, this enables us to make simulation without relying on real-time evolution, thereby mitigating entanglement-related bottlenecks.
Moreover, given that low-energy excited states contribute more significantly and can be resolved with higher accuracy, we anticipate that our method has the potential to outperform existing finite-temperature approaches when operating under the same bond dimension constraint.
It is then beneficial to perform a systematic comparison of our approach against purification and METTS, as originally suggested in Ref.~\onlinecite{Binder2015}. 
Such a study would clarify the regimes in which each method excels and help establish best practices for finite-temperature simulations in strongly correlated quantum systems.
Moreover, we also consider to extend our algorithm to study the contribution from the multiparticle excitation in quantum spin chain~\cite{Vanderstraeten2014, Vanderstraeten2015b}.
We have noticed that a recent development provides a versatile scope for such extension to multi-mode excitation~\cite{Osborne2025}, and we will leave such application to our method as a future work.
Ultimately, we hope to develop a fully-fledged package for simulating low-temperature dynamical properties using GFTNS even in two dimensions, based on our previously proposed eigensolvers using projected entangled-pair states (PEPS)~\cite{Tu2024, peps-excitation}.


\section{Method}
\label{sec:method}

In this study, our target is to investigate the thermal effect for quantum spin chain. 
Our principle ansatz is MPS, which is a compressed representation of quantum many-body state and is known for its efficiency in obtaining ground states in 1D~\cite{White1992}.
We adopt the periodic boundary condition for extracting the properly defined wave numbers, as well as the translational symmetry in the real space.
Note that the latter constraint can be straightforwardly lifted by simply enlarging the size of unit cell.
Our ansatz for the ground state is then represented by
\begin{equation}
|\psi(A)\rangle = 
\begin{diagram}
\draw (0.5, 1.5) .. controls (0, 1.5) and (0, 2) .. (0.5, 2);
\draw (0.5, 1.5) -- (1, 1.5); \draw[] (1, 2) rectangle (2, 1); \draw (1.5, 1.5) node (X) {$A$};
\draw (2, 1.5) -- (3, 1.5); \draw[] (3, 2) rectangle (4, 1); \draw (3.5, 1.5) node {$A$};
\draw (4, 1.5) -- (5, 1.5); \draw[] (5, 2) rectangle (6, 1); \draw (5.5, 1.5) node {$A$}; \draw (6, 1.5) -- (6.5, 1.5); 
\draw (7, 1.5) node {$\ldots$};
\draw (7.5, 1.5) -- (8, 1.5); \draw[] (8, 2) rectangle (9, 1); \draw (8.5, 1.5) node {$A$};
\draw (9, 1.5) -- (10, 1.5); \draw[] (10, 2) rectangle (11, 1); \draw (10.5, 1.5) node {$A$};
\draw (11, 1.5) -- (11.5, 1.5);
\draw (11.5, 1.5) .. controls (12, 1.5) and (12, 2) .. (11.5, 2);
\draw (1.5, 1) -- (1.5, 0.5); \draw (1.5, 0) node {$s_0$};
\draw (3.5, 1) -- (3.5, 0.5); \draw (3.5, 0) node {$s_1$};
\draw (5.5, 1) -- (5.5, 0.5); \draw (5.5, 0) node {$s_2$};
\draw (7, 0) node {$\ldots$};
\draw (8.5, 1) -- (8.5, 0.5); \draw (8.5, 0) node {$s_{N-2}$};
\draw (10.5, 1) -- (10.5, 0.5); \draw (10.5, 0) node {$s_{N-1}$};
\end{diagram},
\label{eq:uniformMPS}
\end{equation}
where $A$ denotes a rank-3 tensor with a local Hilbert space of dimension $d$ and an auxiliary bond dimenssion $\chi$. Size of the latter parameter controls the accuracy of the approximation. 
We adopt the direct variational optimization where the energy gradient can be calculated through the automatic differentiation~\cite{Liao2019}.
Note that due to the antiferromagnetic nature for $S=1/2$ Heisenberg model, we conduct a unitary rotation on the odd sites so that the ground state can be expressed by only one tensor~\cite{Hasik2021}.

Based on the ground state, we then construct excited states using the single-mode approximation~\cite{Haegeman2013a}, which corresponds to a one-particle excitation and works well in various models~\cite{Haegeman2012,Zou2018,Vanderstraeten2018}. The excitation ansatz takes the following form:
\begin{equation}
|\phi_k(B)\rangle=\sum_{j=0}^{N-1}\mathrm{e}^{-ikr_j}\hat{T}^j
\begin{diagram}
\draw (0.5, 1.5) .. controls (0, 1.5) and (0, 2) .. (0.5, 2);
\draw (0.5, 1.5) -- (1, 1.5); \draw[] (1, 2) rectangle (2, 1); \draw (1.5, 1.5) node (X) {$B$};
\draw (2, 1.5) -- (3, 1.5); \draw[] (3, 2) rectangle (4, 1); \draw (3.5, 1.5) node {$A$}; \draw (4, 1.5) -- (4.5, 1.5); 
\draw (5, 1.5) node {$\ldots$};
\draw (5.5, 1.5) -- (6, 1.5); \draw[] (6, 2) rectangle (7, 1); \draw (6.5, 1.5) node {$A$};
\draw (7, 1.5) -- (7.5, 1.5);
\draw (7.5, 1.5) .. controls (8, 1.5) and (8, 2) .. (7.5, 2);
\draw (1.5, 1) -- (1.5, 0.5); \draw (1.5, 0) node {$s_0$};
\draw (3.5, 1) -- (3.5, 0.5); \draw (3.5, 0) node {$s_1$};
\draw (5, 0) node {$\ldots$};
\draw (6.5, 1) -- (6.5, 0.5); \draw (6.5, 0) node {$s_{N-1}$};
\end{diagram},
\label{eq:excitation}
\end{equation}
where the translation operator $\hat{T}$ is defined by $\hat{T}|s_1,s_2,\ldots, s_N\rangle=|s_N,s_1,\ldots, s_{N-1}\rangle$ and satisfies $\hat{T}^N=1$. 
$B$ tensor marks the site of excitation and the wave number $k$ takes values $k = 2\pi m/N$ with $m = 0, 1, \cdots ,N-1$. The index $j$ labels the site for $B$ tensor, and $e^{ik}$ serves as the eigenvalue of the translation operator $\hat{T}$.

While in principle the $B$ tensor can be solved by an iterative eigensolver, here we adopt an alternative way and diagonalize an generalized eigenvalue problem for it
\begin{equation}
  \bm{H}_{\mu\nu}\bm{B}^\nu = E\bm{N}_{\mu\nu}\bm{B}^\nu,
  \label{eq:eigen_prob_for_excited}
\end{equation}
where
\begin{equation}
   \bm{H}_{\mu\nu} = \frac{\partial^2}{\partial\bar{\bm{B}^\mu}\partial \bm{B}^\nu} \bra{\phi_k(\bar{B})} \hat{H} \ket{\phi_k(B)}
   \label{eq:ham_for_excited}
\end{equation}
is the effective Hamiltonian. And
\begin{equation}
   \bm{N}_{\mu\nu} =\frac{\partial^2}{\partial\bar{\bm{B}^\mu}\partial \bm{B}^\nu} \braket{\phi_k(\bar{B})|\phi_k(B)}
   \label{eq:norm_for_excited}
\end{equation}
denotes the effective norm. The indices $\mu$ and $\nu$ label the degrees of freedom associated with $B$ and $\bar{B}$, the complex conjugate of $B$. Their dimensionality matches the size of the Hilbert space containing all accessible quantum states.
As explained in Section~\ref{sec:MPS_Gibbs_state}, Eqs.~(\ref{eq:ham_for_excited}) and (\ref{eq:norm_for_excited}) can be calculated by taking derivative of $\langle G_{\phi}(\lambda', \bar{B})|G_{\phi}(\lambda, B)\rangle$ to $B$ and $\bar{B}$ under $\lambda'$, $\lambda=1$ and $B$, $\bar{B}=0$.
Again thanks to the translational invariance, the summation and thus the derivative can be reduced by one time, saving the computational effort.
The computational cost in the construction of the effective Hamiltonian is $\mathcal{O}(NdD^2\chi^5)$, where $D$ is the bond dimension of MPO for the Hamiltonian.



\par\noindent\emph{\textbf{Acknowledgments}} ---
We acknowledge insightful discussions with Chia-Min Chung, Naoki Kawashima, and Garnet Kin-Lic Chan.
The authors also acknowledge Liusuo Wu for providing us with experimental data.
Numerical calculation was mainly performed at the Supercomputer Center of Institute for Solid State Physics~(ISSP), the University of Tokyo.
W.-L.T. is supported by the Center of Innovation for Sustainable Quantum AI (JST Grant Number JPMJPF2221) and JSPS KAKENHI Grant Number JP25H01545.
J.-Y.C. is supported by National Natural Science Foundation of China (Grants No.~12447107, No.~12304186), Guangdong Basic and Applied Basic Research Foundation (Grant No.~2024A1515013065), and Quantum Science and Technology - National Science and Technology Major Project (Grant No.~2021ZD0302100).
Y.N. acknowledges the financial support by JSPS KAKENHI (Grant Numbers JP23H04869, JP23K03307, and JP25H01506) and MEXT as ``Program for Promoting Researches on the Supercomputer Fugaku'' (Project ID: JPMXP1020230411).



\bibliography{draft}

\end{document}